\documentclass[
reprint,
superscriptaddress,
showpacs,
showkeys,
amsmath,
amssymb,
aps,
prl
]{revtex4-1}

\usepackage{graphicx}
\usepackage{dcolumn}
\usepackage{bm}

\def\babar{\mbox{\sl B\hspace{-0.4em} {\small\sl A}\hspace{-0.37em}
    \sl B\hspace{-0.4em} {\small\sl A\hspace {-0.02em}R}\hspace {+0.2em} }}

\begin{document}

\title{Observation of $D^0$-$\bar{D}^0$ Mixing using the CDF~II
  Detector}

\affiliation{Institute of Physics, Academia Sinica, Taipei, Taiwan 11529, Republic of China}
\affiliation{Argonne National Laboratory, Argonne, Illinois 60439, USA}
\affiliation{University of Athens, 157 71 Athens, Greece}
\affiliation{Institut de Fisica d'Altes Energies, ICREA, Universitat Autonoma de Barcelona, E-08193, Bellaterra (Barcelona), Spain}
\affiliation{Baylor University, Waco, Texas 76798, USA}
\affiliation{Istituto Nazionale di Fisica Nucleare Bologna, \ensuremath{^{jj}}University of Bologna, I-40127 Bologna, Italy}
\affiliation{University of California, Davis, Davis, California 95616, USA}
\affiliation{University of California, Los Angeles, Los Angeles, California 90024, USA}
\affiliation{Instituto de Fisica de Cantabria, CSIC-University of Cantabria, 39005 Santander, Spain}
\affiliation{Carnegie Mellon University, Pittsburgh, Pennsylvania 15213, USA}
\affiliation{Enrico Fermi Institute, University of Chicago, Chicago, Illinois 60637, USA}
\affiliation{Comenius University, 842 48 Bratislava, Slovakia; Institute of Experimental Physics, 040 01 Kosice, Slovakia}
\affiliation{Joint Institute for Nuclear Research, RU-141980 Dubna, Russia}
\affiliation{Duke University, Durham, North Carolina 27708, USA}
\affiliation{Fermi National Accelerator Laboratory, Batavia, Illinois 60510, USA}
\affiliation{University of Florida, Gainesville, Florida 32611, USA}
\affiliation{Laboratori Nazionali di Frascati, Istituto Nazionale di Fisica Nucleare, I-00044 Frascati, Italy}
\affiliation{University of Geneva, CH-1211 Geneva 4, Switzerland}
\affiliation{Glasgow University, Glasgow G12 8QQ, United Kingdom}
\affiliation{Harvard University, Cambridge, Massachusetts 02138, USA}
\affiliation{Division of High Energy Physics, Department of Physics, University of Helsinki, FIN-00014, Helsinki, Finland; Helsinki Institute of Physics, FIN-00014, Helsinki, Finland}
\affiliation{University of Illinois, Urbana, Illinois 61801, USA}
\affiliation{The Johns Hopkins University, Baltimore, Maryland 21218, USA}
\affiliation{Institut f\"{u}r Experimentelle Kernphysik, Karlsruhe Institute of Technology, D-76131 Karlsruhe, Germany}
\affiliation{Center for High Energy Physics: Kyungpook National University, Daegu 702-701, Korea; Seoul National University, Seoul 151-742, Korea; Sungkyunkwan University, Suwon 440-746, Korea; Korea Institute of Science and Technology Information, Daejeon 305-806, Korea; Chonnam National University, Gwangju 500-757, Korea; Chonbuk National University, Jeonju 561-756, Korea; Ewha Womans University, Seoul, 120-750, Korea}
\affiliation{Ernest Orlando Lawrence Berkeley National Laboratory, Berkeley, California 94720, USA}
\affiliation{University of Liverpool, Liverpool L69 7ZE, United Kingdom}
\affiliation{University College London, London WC1E 6BT, United Kingdom}
\affiliation{Centro de Investigaciones Energeticas Medioambientales y Tecnologicas, E-28040 Madrid, Spain}
\affiliation{Massachusetts Institute of Technology, Cambridge, Massachusetts 02139, USA}
\affiliation{University of Michigan, Ann Arbor, Michigan 48109, USA}
\affiliation{Michigan State University, East Lansing, Michigan 48824, USA}
\affiliation{Institution for Theoretical and Experimental Physics, ITEP, Moscow 117259, Russia}
\affiliation{University of New Mexico, Albuquerque, New Mexico 87131, USA}
\affiliation{The Ohio State University, Columbus, Ohio 43210, USA}
\affiliation{Okayama University, Okayama 700-8530, Japan}
\affiliation{Osaka City University, Osaka 558-8585, Japan}
\affiliation{University of Oxford, Oxford OX1 3RH, United Kingdom}
\affiliation{Istituto Nazionale di Fisica Nucleare, Sezione di Padova, \ensuremath{^{kk}}University of Padova, I-35131 Padova, Italy}
\affiliation{University of Pennsylvania, Philadelphia, Pennsylvania 19104, USA}
\affiliation{Istituto Nazionale di Fisica Nucleare Pisa, \ensuremath{^{ll}}University of Pisa, \ensuremath{^{mm}}University of Siena, \ensuremath{^{nn}}Scuola Normale Superiore, I-56127 Pisa, Italy, \ensuremath{^{oo}}INFN Pavia, I-27100 Pavia, Italy, \ensuremath{^{pp}}University of Pavia, I-27100 Pavia, Italy}
\affiliation{University of Pittsburgh, Pittsburgh, Pennsylvania 15260, USA}
\affiliation{Purdue University, West Lafayette, Indiana 47907, USA}
\affiliation{University of Rochester, Rochester, New York 14627, USA}
\affiliation{The Rockefeller University, New York, New York 10065, USA}
\affiliation{Istituto Nazionale di Fisica Nucleare, Sezione di Roma 1, \ensuremath{^{qq}}Sapienza Universit\`{a} di Roma, I-00185 Roma, Italy}
\affiliation{Mitchell Institute for Fundamental Physics and Astronomy, Texas A\&M University, College Station, Texas 77843, USA}
\affiliation{Istituto Nazionale di Fisica Nucleare Trieste, \ensuremath{^{rr}}Gruppo Collegato di Udine, \ensuremath{^{ss}}University of Udine, I-33100 Udine, Italy, \ensuremath{^{tt}}University of Trieste, I-34127 Trieste, Italy}
\affiliation{University of Tsukuba, Tsukuba, Ibaraki 305, Japan}
\affiliation{Tufts University, Medford, Massachusetts 02155, USA}
\affiliation{University of Virginia, Charlottesville, Virginia 22906, USA}
\affiliation{Waseda University, Tokyo 169, Japan}
\affiliation{Wayne State University, Detroit, Michigan 48201, USA}
\affiliation{University of Wisconsin, Madison, Wisconsin 53706, USA}
\affiliation{Yale University, New Haven, Connecticut 06520, USA}

\author{T.~Aaltonen}
\affiliation{Division of High Energy Physics, Department of Physics, University of Helsinki, FIN-00014, Helsinki, Finland; Helsinki Institute of Physics, FIN-00014, Helsinki, Finland}
\author{S.~Amerio\ensuremath{^{kk}}}
\affiliation{Istituto Nazionale di Fisica Nucleare, Sezione di Padova, \ensuremath{^{kk}}University of Padova, I-35131 Padova, Italy}
\author{D.~Amidei}
\affiliation{University of Michigan, Ann Arbor, Michigan 48109, USA}
\author{A.~Anastassov\ensuremath{^{w}}}
\affiliation{Fermi National Accelerator Laboratory, Batavia, Illinois 60510, USA}
\author{A.~Annovi}
\affiliation{Laboratori Nazionali di Frascati, Istituto Nazionale di Fisica Nucleare, I-00044 Frascati, Italy}
\author{J.~Antos}
\affiliation{Comenius University, 842 48 Bratislava, Slovakia; Institute of Experimental Physics, 040 01 Kosice, Slovakia}
\author{G.~Apollinari}
\affiliation{Fermi National Accelerator Laboratory, Batavia, Illinois 60510, USA}
\author{J.A.~Appel}
\affiliation{Fermi National Accelerator Laboratory, Batavia, Illinois 60510, USA}
\author{T.~Arisawa}
\affiliation{Waseda University, Tokyo 169, Japan}
\author{A.~Artikov}
\affiliation{Joint Institute for Nuclear Research, RU-141980 Dubna, Russia}
% ...

\collaboration{CDF Collaboration}
\altaffiliation[With visitors from]{
\ensuremath{^{a}}University of British Columbia, Vancouver, BC V6T 1Z1, Canada,
\ensuremath{^{b}}Istituto Nazionale di Fisica Nucleare, Sezione di Cagliari, 09042 Monserrato (Cagliari), Italy,
\ensuremath{^{c}}University of California Irvine, Irvine, CA 92697, USA,
\ensuremath{^{d}}Institute of Physics, Academy of Sciences of the Czech Republic, 182~21, Czech Republic,
\ensuremath{^{e}}CERN, CH-1211 Geneva, Switzerland,
\ensuremath{^{f}}Cornell University, Ithaca, NY 14853, USA,
\ensuremath{^{g}}University of Cyprus, Nicosia CY-1678, Cyprus,
\ensuremath{^{h}}Office of Science, U.S. Department of Energy, Washington, DC 20585, USA,
\ensuremath{^{i}}University College Dublin, Dublin 4, Ireland,
\ensuremath{^{j}}ETH, 8092 Z\"{u}rich, Switzerland,
\ensuremath{^{k}}University of Fukui, Fukui City, Fukui Prefecture, Japan 910-0017,
\ensuremath{^{l}}Universidad Iberoamericana, Lomas de Santa Fe, M\'{e}xico, C.P. 01219, Distrito Federal,
\ensuremath{^{m}}University of Iowa, Iowa City, IA 52242, USA,
\ensuremath{^{n}}Kinki University, Higashi-Osaka City, Japan 577-8502,
\ensuremath{^{o}}Kansas State University, Manhattan, KS 66506, USA,
\ensuremath{^{p}}Brookhaven National Laboratory, Upton, NY 11973, USA,
\ensuremath{^{q}}University of Manchester, Manchester M13 9PL, United Kingdom,
\ensuremath{^{r}}Queen Mary, University of London, London, E1 4NS, United Kingdom,
\ensuremath{^{s}}University of Melbourne, Victoria 3010, Australia,
\ensuremath{^{t}}Muons, Inc., Batavia, IL 60510, USA,
\ensuremath{^{u}}Nagasaki Institute of Applied Science, Nagasaki 851-0193, Japan,
\ensuremath{^{v}}National Research Nuclear University, Moscow 115409, Russia,
\ensuremath{^{w}}Northwestern University, Evanston, IL 60208, USA,
\ensuremath{^{x}}University of Notre Dame, Notre Dame, IN 46556, USA,
\ensuremath{^{y}}Universidad de Oviedo, E-33007 Oviedo, Spain,
\ensuremath{^{z}}CNRS-IN2P3, Paris, F-75205 France,
\ensuremath{^{aa}}Universidad Tecnica Federico Santa Maria, 110v Valparaiso, Chile,
\ensuremath{^{bb}}The University of Jordan, Amman 11942, Jordan,
\ensuremath{^{cc}}Universite catholique de Louvain, 1348 Louvain-La-Neuve, Belgium,
\ensuremath{^{dd}}University of Z\"{u}rich, 8006 Z\"{u}rich, Switzerland,
\ensuremath{^{ee}}Massachusetts General Hospital, Boston, MA 02114 USA,
\ensuremath{^{ff}}Harvard Medical School, Boston, MA 02114 USA,
\ensuremath{^{gg}}Hampton University, Hampton, VA 23668, USA,
\ensuremath{^{hh}}Los Alamos National Laboratory, Los Alamos, NM 87544, USA,
\ensuremath{^{ii}}Universit\`{a} degli Studi di Napoli Federico I, I-80138 Napoli, Italy
}
\noaffiliation

\date{November 1, 2013}

\begin{abstract}

  We measure the time dependence of the ratio of decay rates for
  $D^0 \rightarrow K^+\pi^-$ to the Cabibbo-favored decay
  $D^0 \rightarrow K^-\pi^+$.  The charge conjugate decays are
  included. A signal of $3.3 \times 10^4$ $D^{*+} \to \pi^+ D^0$, $D^0
  \rightarrow K^+\pi^-$ decays is obtained with $D^0$ proper decay
  times between 0.75 and 10 mean $D^0$ lifetimes.  The data were
  recorded with the CDF~II detector at the Fermilab Tevatron and
  correspond to an integrated luminosity of 9.6 fb$^{-1}$ for
  $p\bar{p}$ collisions at $\sqrt{s}$ = 1.96 TeV.  Assuming {\it CP}
  conservation, we search for $D^0$-$\bar{D}^0$ mixing and measure
  the mixing parameters to be $R_D = (3.51 \pm 0.35) \times 10^{-3}$,
  $y' = (4.3 \pm 4.3) \times 10^{-3}$, and $x'^2 = (0.08 \pm 0.18)
  \times 10^{-3}$.  We report Bayesian probability intervals in the
  $x'^2$-$y'$ plane and find that the significance of excluding the
  no-mixing hypothesis is equivalent to 6.1 Gaussian standard deviations,
  providing the second observation of $D^0$-$\bar{D}^0$ mixing from a
  single experiment.

\end{abstract}




\maketitle

A neutral meson that is a superposition of weakly decaying mass
eigenstates can spontaneously change into its antiparticle.  This
process is referred to as mixing and is well established for
$K^0$, $B^0$, and $B^0_s$ mesons \cite{ref:PDG}. The mixing of these
mesons is understood within the framework of the
standard model with the Cabibbo-Kobayashi-Maskawa quark-mixing matrix.
Substantial evidence exists for $D^0$-$\bar{D}^0$ mixing
\cite{ref:Asner, ref:BaBar_mixing, ref:Belle_mixing, ref:CDF_mixing,
  ref:BaBar_mixing_lifetime_diff}, and the process was recently
observed in the $K\pi$ channel by the LHCb experiment \cite{ref:LHCb}.
In the standard model, $D^0$-$\bar{D}^0$ mixing is a weak-interaction
process that occurs primarily through long-range virtual
intermediate states that consist of common decay channels for particle
and antiparticle, such as $\pi^+\pi^-$.
The prediction of the mixing rate has significant uncertainty because
it requires a strong-interaction model. A direct calculation from
quantum chromodynamics is not currently possible ~\cite{ref:cicerone,
ref:FGLNP-2004}.
Mixing could also result from
exotic particles that appear as virtual states in a short-range box
diagram. Such a process could provide indirect evidence for physics
beyond the standard model \cite{ref:Ciuchini, ref:GHPP-2007}.  It is
of great interest to verify independently the observation of
$D^0$-$\bar{D}^0$ mixing in a single experiment and to improve the
measurement of the mixing parameters.  We report a measurement using
the decay $D^0 \to K^+\pi^-$ and its charge conjugate.

The decay $D^0 \to K^+\pi^-$ can arise from mixing of a $D^0$ state to
a $\bar{D}^0$ state, followed by a Cabibbo-favored (CF) decay, or from
a doubly Cabibbo-suppressed (DCS) decay of a $D^0$.  (In this
Letter, reference to a specific decay chain implicitly includes the
charge-conjugate decay.)  The mixing measurement is based on the ratio
$R$ of $D^0 \to K^+\pi^-$ to $D^0 \to K^-\pi^+$ decay rates.  This
ratio can be approximated \cite{ref:Blaylock-mix, ref:cleo} as a
quadratic function of $t/\tau$, where $t$ is the proper decay time and
$\tau$ is the mean $D^0$ lifetime:  
\begin{equation}
R(t/\tau) = R_D + \sqrt{R_D} {y}^{\prime} \,(t/\tau) + \frac{ x^{\prime 2} + 
         y^{\prime 2} }{4} \, (t/\tau)^2.
\label{eqn:R(t)}
\end{equation}
This form is valid under the assumption of {\it CP} conservation and
small values for the parameters $x = \Delta m / \Gamma$ and $y =
\Delta \Gamma / 2 \Gamma$, where $\Delta m$ is the mass difference
between the mass eigenstates, $\Delta \Gamma$ is the decay-width
difference, and $\Gamma$ is the mean decay width of the eigenstates.
The parameter $R_D$ is the squared modulus of the ratio of DCS to CF
amplitudes.  The parameters $x^{\prime}$ and $y^{\prime}$ are linear
combinations of $x$ and $y$ according to the relations
$$
x^{\prime} = x ~{\rm cos}~\delta + y ~{\rm sin} ~\delta ~~~{\rm and}~~~ 
y^{\prime} = - x ~{\rm sin} ~\delta + y ~{\rm cos} ~\delta ,
$$
where $\delta$ is the strong-interaction phase difference between the
DCS and CF amplitudes.  In the absence of mixing, $x' = y' = 0$ and
$R(t/\tau) = R_D$. 

The measurement uses the full data set collected by the CDF~II
detector at the Fermilab Tevatron collider, from February 2002 to
September 2011, corresponding to an integrated luminosity of
$9.6$~fb$^{-1}$ for $p\bar{p}$ collisions at $\sqrt{s}=1.96$~TeV.  We
previously reported evidence for $D^0$-$\bar{D}^0$ mixing
\cite{ref:CDF_mixing} based on a subset of the data corresponding to
an integrated luminosity of 1.5 fb$^{-1}$.  The multipurpose CDF~II
detector \cite{ref:CDF} is a magnetic spectrometer surrounded by a
calorimeter and a muon detector.  The detector components pertinent to
this analysis are the silicon microstrip vertex detector, the
multi-wire drift chamber, and the 1.4~T magnet, which together measure
the trajectories and momenta of charged particles.  The drift chamber
measures ionization energy loss for charged particles, which is used
for particle identification.  Events are selected in real time with a
trigger system developed for a broad class of heavy-flavor decays.
The trigger \cite{ref:CDF-RB} selects events with a pair of
oppositely charged particles originating from a decay point separated
by at least 200 $\mu$m from the beamline in the transverse plane.

We identify the CF decay $D^0 \to K^- \pi^+$ through the right-sign
(RS) decay chain $D^{*+} \to \pi^+ D^0$, $D^0 \to K^- \pi^+$.  The
decay $D^0 \to K^+ \pi^-$ is identified through the wrong-sign (WS)
decay chain $D^{*+} \to \pi^+ D^0$, $D^0 \to K^+ \pi^-$.  The relative
charges of the pions determine whether the decay chain is RS (same
charge) or WS (opposite charge).  The RS and WS $D^*$ decays have the
same kinematic distributions, and may differ only in decay time
distributions.  To reduce systematic uncertainties, we use the same
selection criteria (cuts) for both the RS and WS decay modes.
Analysis cuts were optimized before the WS candidates were revealed,
and were chosen to maximize the expected WS signal significance.

The $D^0$ candidate reconstruction starts with a pair of tracks from
oppositely-charged particles that satisfy the trigger requirements.
The tracks are considered with both $K^-\pi^+$ and $\pi^-K^+$
interpretations.  A third track, required to have transverse momentum
in the range [0.4, 2.0] GeV/$c$ (see \cite{ref:CDF_Dhh}), is used to
form a $D^*$ candidate when considered as a pion and combined with the
$D^0$ candidate.

We use a method primarily based on $D^0$ decay kinematics to reduce
the background to the WS signal from RS decays where the $D^0$ decay
tracks are misidentified because the kaon and pion assignments are
mistakenly interchanged.  As determined from data for RS $D^*$'s,
96\% of $D^0$ decays with correct mass assignments are reconstructed
with invariant masses within 20 MeV/$c^2$ of the $D^0$ mass.  The
invariant mass distribution for misidentified $D^0$ decays is much
broader, and has only 23\% of the events within the same mass range.
We remove WS candidates that have a RS mass within that range.  To
further reject $D^*$ candidates with misidentified decay tracks, we
impose a cut based on particle identification, described in
Ref.~\cite{ref:CDF-RB}, that is used to choose between $K^-\pi^+$ and
$\pi^-K^+$ assignments for the $D^0$ decay tracks.
 
We use a series of cuts based on the decay topology of signal events
in which the $D^{*}$ is produced at the collision point and the $D^0$
travels a measurable distance before it decays.  The topological cuts
reduce background from combinations involving one or more tracks that
do not originate from the $D^*$ decay chain of interest.  We require
the significance of the transverse decay length to satisfy
$L_{xy}/\sigma_{xy}>4$, where $L_{xy}$ is the distance between the
collision point (measured on an event-by-event basis) and the
reconstructed $D^0$ decay point in the plane transverse to the
beamline, and $\sigma_{xy}$ is the uncertainty on $L_{xy}$.  The
transverse impact parameter $d_0$ is the distance of closest approach
in the transverse plane between a track (or reconstructed particle)
and the collision point.  The $D^{*}$-decay pion must have
$d_0~<~600~\mu$m, and it must also have a distance of closest approach to
the collision point less than 1.5 cm along the beamline.  To reduce
the contribution of nonprompt $D^*$'s produced in beauty-particle
decays, we require $d_0~<~60~\mu$m for the inferred track of the $D^0$
candidate. The remaining contribution of nonprompt
$D^*$ mesons is taken into account in the analysis of the time
dependence of the WS/RS ratio, as discussed later.

The ratio $t/\tau$ is determined for each $D^0$ candidate by $t/\tau =
M_{D^0} L_{xy} / (p_T \tau)$, where $M_{D^0} = 1.8648$~GeV/$c^2$ and
$\tau$ = 410.1~fs are the world-average values for the $D^0$ mass and
lifetime, respectively \cite{ref:PDG}.  To study $R(t/\tau)$, we
divide the data into 20 bins of $t/\tau$ ranging from 0.75 to 10.0,
choosing bins of increasing size from 0.25 to 2.0 to reduce
statistical uncertainty per bin at larger times.  The bin sizes are
larger than the typical $t/\tau$ resolution of 0.16.

\begin{figure}[hbpt]
\begin{center}
  \includegraphics[width=2.7in]{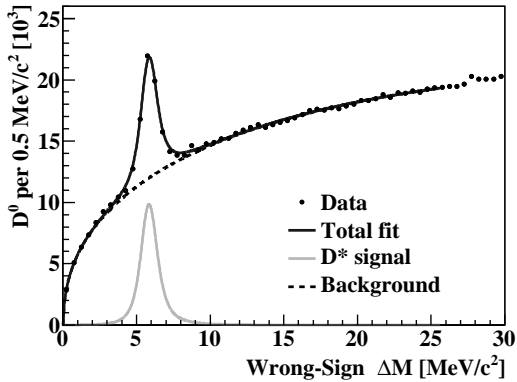}
  \caption{Time-integrated distribution for wrong-sign $D^0
    \rightarrow K^+ \pi^- $ signal yield as a function of $\Delta M$.
    The signal yield for each bin of $\Delta M$ is determined from a
    fit to the corresponding $M_{K\pi}$ distribution.  The result of a
    $\chi^2$ fit to the $\Delta M$ distribution is shown.}
\label{fig:WS_mass_diff}
\end{center}
\end{figure}

The RS and WS candidates in each $t/\tau$ bin are further divided into
60 bins of mass difference $\Delta M$ $\equiv$ $M_{K^+\pi^-\pi^+} -
M_{K^+\pi^-} - M_{\pi^{+}}$ for WS candidates, and analogously for RS
candidates. For each of the 1200 WS and 1200 RS $\Delta M$ bins, the
$D^0$ signal yield is determined from a fit to the corresponding
distribution of $M_{K\pi}$, which has 60 bins in the range of 1.80 - 1.92~GeV/$c^2$.
For the $M_{K\pi}$ fit, the signal shape is
modeled by a double-Gaussian form with a low-mass tail, and the
combinatorial background is modeled by an exponential.  For the WS
$M_{K\pi}$ fit, a Gaussian term models the RS misidentified background,
with mean and width determined from the data.  The $D^*$ signal yield for each
time bin is determined from a $\chi^2$ fit of the $D^0$ signal
yield versus $\Delta M$.  The signal shape is modeled by a double
Gaussian and an asymmetric tail.  The background shape is modeled by
the product of a power law and an exponential function.  The WS
signal shape parameters for both the $M_{K\pi}$ and $\Delta M$
distributions are fixed to the RS parameters.  For each $M_{K\pi}$ and
$\Delta M$ distribution, the parameters for the background shape are
allowed to float.  The amplitudes of the signal and background are
determined independently for all $M_{K\pi}$ and $\Delta M$ fits.  The
RS distributions have similar absolute amounts of background as the WS
distributions, but the RS signal is about 230 times larger.  A
detailed description of the functional forms for the signal and
background shapes is presented in Ref.~\cite{ref:CDF_Dhh}.
The $M_{K\pi}$ distributions from a subset of the data, which are
characteristic of the full data set, are reported in
\cite{ref:CDF_mixing,ref:CDF-RB}.
The 2400 $M_{K\pi}$ histograms are well fit by
the functional forms used with 57 degrees of freedom.
The distribution of $\chi^2$/dof has a mean
of 1.2 and a standard deviation of 0.4.  The time-integrated WS
$\Delta M$ distribution is shown in Fig.~\ref{fig:WS_mass_diff}.  The
time-integrated signal yields are $(3.27 \pm 0.04) \times 10^{4}$
(WS) and $(7.604 \pm 0.005) \times 10^{6}$ (RS).

\begin{figure}[htbp]
\begin{center}
\includegraphics[width=2.7in]{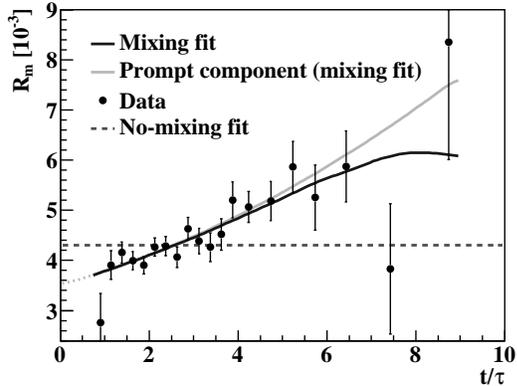}
\caption{
Measured ratio $R_m$ of wrong-sign to right-sign $D^*$ decays as 
a function of normalized proper decay time. The results of
a least-squares fit are shown for $R_m^{pred}$ (mixing fit) and
the prompt component $R$. A fit assuming no mixing is
clearly incompatible with the data.}
\label{fig:ratio_data}
\end{center}
\end{figure}

The measured ratio $R_m$ of WS to RS signal for each of the 20
$t/\tau$ bins is shown in Fig.~\ref{fig:ratio_data}. The data point
for each bin is located at the mean value of $t/\tau$ for the RS
signal in that bin.  Each error bar is determined from the fit
uncertainties of the WS and RS signal yields.  The large uncertainty
in the smallest $t/\tau$ bin is due to the small event yield caused by
the trigger turn-on.  After the trigger turn-on, the uncertainties
increase with $t/\tau$ because of the exponential fall-off in the
number of decays.

We assume that the WS/RS ratio of decay rates $R(t/\tau)$, given by
Eq.~(\ref{eqn:R(t)}), is the same for $D^0$'s produced in
beauty-particle decays as for promptly-produced $D^0$'s, but with
decay time measured from the beauty-particle decay point.  The
predicted value of $R_m$ for a given $t/\tau$ bin can be expressed in
terms of contributions from prompt and nonprompt production according
to
\begin{equation}
R_m^{pred} 
= R(t/\tau) \left[ 1 - f_B(t/\tau)\right] + R_B(t/\tau) f_B(t/\tau), 
\label{eqn:R_expected}
\end{equation}
where
$f_B(t/\tau)$ is the fraction of nonprompt RS
$D^*$ decays and $R_B(t/\tau)$ is the WS/RS ratio of nonprompt $D^*$
decays.  For nonprompt decays, the measured decay time is due to the
combination of the decay times for the beauty-particle parent and its
$D^0$ daughter.  The function $f_B(t/\tau)$ is determined from data,
and $R_B(t/\tau)$ is determined from a full detector simulation.

\begin{figure}[hbpt]
\begin{center}
\includegraphics[width=2.7in]{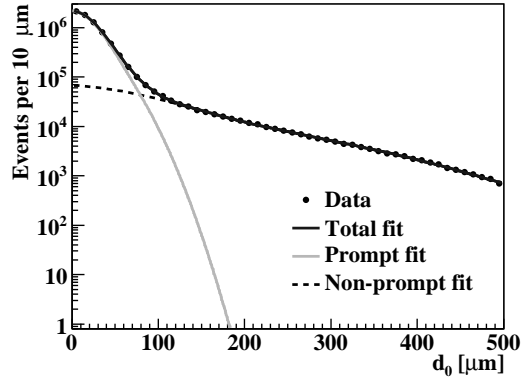}
\caption{
Distribution of transverse impact parameter $d_0$ for 
right-sign $D^0$ candidates from all $t/\tau$ bins.
The narrow peak is due to promptly-produced $D^0$ mesons
and the broad distribution is due to nonprompt $D^0$ mesons from 
beauty-particle decay.  
For candidates with $d_0 < 60 ~\mu$m, 95\%
are produced promptly at the collision point and 5\% are nonprompt.}
\label{fig:ip_dist}
\end{center}
\end{figure}

The function $f_B(t)$ is determined from the $d_0$ distribution of $D^0$'s
from RS $D^*$ decays, as illustrated in Fig.~\ref{fig:ip_dist}.  For each bin
of $t/\tau$, the $d_0$ distribution is obtained by selecting RS events
with $ 4 < \Delta M < 8$ MeV/$c^{2}$ and 1.848 $< M_{K\pi} < 1.880$
GeV/$c^{2}$ ($\pm$2$\sigma$) and subtracting background determined
from the $M_{K\pi}$ sidebands (low-mass 1.808--1.824 GeV/$c^{2}$, high-mass
1.904--1.920 GeV/$c^{2}$). The peak at small $d_0$ is due to the prompt
component.  The broad distribution extending to large $d_0$ is due to
the nonprompt component.  The prompt and nonprompt components are
each modeled with the sum of two Gaussians. The fraction $f_B$ is
determined in each time bin for the region $d_0 < 60 ~\mu$m, which is
dominated by the prompt component.  The time dependence of $f_B$ is
characterized by a five-parameter polynomial fit to the values from
each time bin.  The value of $f_B$ is (1.5$\pm$0.4)\% at $t/\tau =
1.4$ and increases with $t/\tau$ due to the faster decay rate of $D^0$
mesons compared to beauty particles.  At $t/\tau = 6.4$, $f_B$ =
$(24\pm 1)$\%.
The function $R_B(t/\tau)$ can be expressed in terms of a function
$H(t/\tau, t'/\tau)$ which gives the distribution of nonprompt $D^0$
decays versus $t/\tau$ for a given $t'/\tau$, where $t'/\tau$ is
measured from the decay point of the beauty particle.  The function
$H$ is determined from a full detector simulation of beauty particle
to $D^*$ decays for the 20 bins of $t/\tau$, and 100 bins of
$t^{\prime}/\tau$.  The function $R_B(t/\tau)$ is given by
\begin{equation}
  R_B(t_i/\tau) = 
  \frac{\sum_{j=1}^{100} H(t_i/\tau, t'_j/\tau) R(t'_j/\tau)}{\sum_{j=1}^{100} 
    H(t_i/\tau, t'_j/\tau)},
\end{equation}
where $i$ and $j$ denote the bins in $t$ and $t'$.  Note that $R_B$
depends directly on the prompt $D^*$ WS/RS ratio $R$ defined in
Eq.~(\ref{eqn:R(t)}).

To fit for the mixing parameters, we define
\begin{equation}
  \chi^2 = \sum_{i=1}^{20} \left[ 
\frac{R_m(t_i/\tau) - R_m^{pred}(t_i/\tau)} {\sigma_i} 
\right]^2 + C_B + C_H,
\label{eqn:chi2_Rm}
\end{equation}
where $\sigma_i$ is the uncertainty on $R_m(t_i/\tau)$. The term $C_B$
are Gaussian constraints on the five fitted parameters describing
$f_B(t/\tau)$, and $C_H$ are Gaussian constraints on the values of
$H(t_i/\tau, t'_j/\tau)$. The statistical uncertainties on the
$H(t_i/\tau, t'_j/\tau)$ are due to the number of events in the
simulation.  The mixing parameters $R_D$, $y'$, and $x'^2$, and the
Gaussian-constrained parameters for $f_B$ and $H$ are found by
minimizing the $\chi^2$ defined in Eq.~(\ref{eqn:chi2_Rm}).

To verify the self-consistency of the analysis procedure, we simulate
distributions of $M_{K\pi}$ and $\Delta M$ for different assumed
values of the mixing parameters $R_D$, $y'$, and $x'^2$. We generate
400 samples for each of four different sets of mixing parameters. For
each parameter set, the distributions of fitted parameters have mean
values consistent with the input parameters.

We examine various sources of systematic uncertainty in the analysis
procedure.  The effect on the WS signal yields due to the uncertainty
in the signal shapes used to fit the $M_{K\pi}$ and $\Delta M$
distributions is studied by independently varying the shape parameters
by $\pm 1 \sigma$.  For each parameter, the resulting variation of
the signal yield is negligible compared to the statistical
uncertainty.  We check the sensitivity of the WS and RS signals to the
assumed shape of the $M_{K\pi}$ background by using simulations with
alternative forms for the background shape.  The alternative forms
include explicit shapes for backgrounds due to $D^+ \rightarrow K^-
\pi^+ \pi^+$ decays, determined from data, and partially reconstructed
charm-particle decays, based on a full detector simulation.  In both
simulation studies, the mixing parameters are found to be consistent with
those generated.  To determine the sensitivity of $R_m$ to $R_B$, we
fit the $d_0$ distributions with an alternative shape function,
leading to an alternate form for $R_B$ with larger values at small
$t/\tau$. The resulting change in $R_m$ is negligible, which can be
understood from the small fraction of nonprompt $D^*$s at small
$t/\tau$. To check the sensitivity of $R_m$ to $H(t/\tau,t'/\tau)$, we
scale $t/\tau$ and $t'/\tau$ by $\pm$10\%.  The resulting changes in
$R_m$ are negligible compared to statistical uncertainties.
 
The results for the mixing parameters are given in Table
\ref{tab:param}.  The parameter values are highly correlated with
correlation coefficients of $-0.97$ for the ($R_D,y'$) term, 0.90 for
$(R_D,x'^2)$, and $-0.98$ for $(y',x'^2)$.  The resulting functions
$R_m^{pred}(t/\tau)$ and $R(t/\tau)$, describing the prompt component,
are shown in Fig.~\ref{fig:ratio_data}. The two functions differ at
large $t/\tau$ due to the effect of nonprompt $D^*$ production.  As
shown in Table \ref{tab:param}, our results for the mixing parameters
are consistent with previous measurements and our measurement of
$x'^2$ is comparable in precision to that from LHCb.

\begin{table}[htbp]
  \caption
  {Mixing parameter results and comparison with previous
    measurements.  All results use $D^0 \rightarrow K^+\pi^-$ decays and
    fits assuming no {\it CP} violation.  The uncertainties include
    statistical and systematic components. The significance for excluding
    the no-mixing hypothesis is given in terms of the equivalent number of
    Gaussian standard deviations $\sigma$.}
\begin{center}
\begin{tabular}{lcccc}
\hline \hline
Expt. & $R_D (10^{-3})$ & $y' \, (10^{-3})$ & $x'^2 \, (10^{-3})$
      & $\sigma$(no mix.) \\
\hline
This meas. & $3.51 \pm 0.35$  & $~~4.3 \pm 4.3$ &  ~~$0.08 \pm 0.18$ & $6.1$ \\
Belle \cite{ref:Belle_dcs}     
& $3.64 \pm 0.17$ & $0.6^{+4.0}_{-3.9}$ & $0.18^{+0.21}_{-0.23}$ & $2.0$ \\
\babar \cite{ref:BaBar_mixing}  
& $3.03 \pm 0.19$ & $~~9.7 \pm 5.4$ & $-0.22 \pm 0.37$ & $3.9$ \\
CDF \cite{ref:CDF_mixing}
& $3.04 \pm 0.55$   & $~~8.5 \pm 7.6$ &  $-0.12 \pm 0.35$  & $3.8$ \\
LHCb \cite{ref:LHCb}                 
&  $3.52 \pm 0.15$  & $~~7.2  \pm 2.4$ & $-0.09 \pm 0.13$ & $9.1$ \\
\hline \hline
\end{tabular}
\end{center}
\label{tab:param}
\end{table}

\begin{figure}[htbp]
\begin{center}
\includegraphics[width=2.7in]{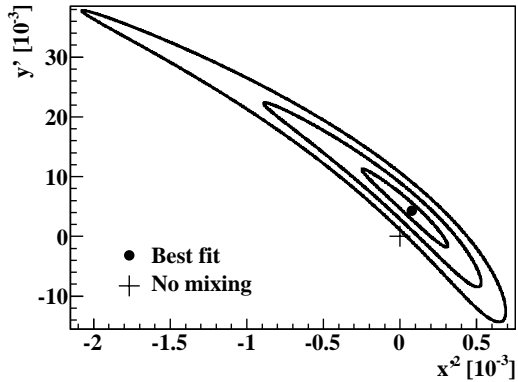}
\caption{Contours in $x'^2 - y'$ parameter space bounding regions with 
Bayesian posterior probability corresponding to 1, 3, and 5 Gaussian standard
deviations.}
\label{fig:contour}
\end{center}
\end{figure}

A fit assuming no mixing, shown in Fig.~\ref{fig:ratio_data}, is
clearly incompatible with the data.  We quantify this incompatibility
using both Bayesian and frequentist methods.  We define a likelihood
${\cal L} = \exp(-\chi^2/2)$, normalized over the mixing parameter
space, with $\chi^2$ as defined in Eq.~(\ref{eqn:chi2_Rm}).  We
compute contours bounding a region with a given value of Bayesian
posterior probability.  A uniform prior is used for $R_D$, $y'$, and
$x'^2$, and $R_D$ is treated as a nuisance parameter.  The contours
are shown in Fig.~\ref{fig:contour}. The no-mixing point, $y' = x'^2 =
0$, lies on the contour corresponding to 6.1 Gaussian (1D) standard
deviations. If the physical restriction $x'^2 > 0$ is imposed, there
is no change in the no-mixing significance.  Using alternative priors,
uniform in $x'$ or $y'^2$, the no-mixing significance is 6.3~$\sigma$.
A frequentist test statistic $\Delta \chi^2$ is formed from the
difference in $\chi^2$ between a fit with $y' = x'^2 = 0$ and a fit
with all three mixing parameters floating. For the data, $\Delta
\chi^2 = 58.75 - 16.91 = 41.84 $.  A frequentist $p$-value is obtained
by simulating $R_m$ distributions for the 20 time bins using $y' = x'^2 = 0$
and evaluating $\Delta \chi^2$ using the same procedure as for data.
In $10^{10}$ samples, 6 are found with $\Delta \chi^2 > 41.8$,
giving a $p$-value corresponding to 6.1~$\sigma$.

In summary, we measure the time dependence of the ratio of decay rates
for $D^0 \rightarrow K^+\pi^-$ to the Cabibbo-favored decay $D^0
\rightarrow K^-\pi^+$. A signal of $3.3 \times 10^4$ $D^{*+} \to \pi^+
D^0$, $D^0 \rightarrow K^+\pi^-$ decays is obtained with proper decay
times between 0.75 and 10 mean $D^0$ lifetimes.  The data sample
recorded with the CDF~II detector at the Fermilab Tevatron corresponds
to an integrated luminosity of 9.6 fb$^{-1}$ for $p\bar{p}$ collisions
at $\sqrt{s}$ = 1.96 TeV.  Assuming {\it CP} conservation, we measure
the $D^0$-$\bar{D}^0$ mixing parameters to be $R_D = (3.51 \pm 0.35)
\times 10^{-3}$, $y' = (4.3 \pm 4.3) \times 10^{-3}$, and $x'^2 =
(0.08 \pm 0.18) \times 10^{-3}$, providing important accuracy for the
world averages.  We report contours in the $x'^2$-$y'$ plane which bound
regions of a given Bayesian posterior probability; 
we find that the significance of excluding the no-mixing hypothesis is
equivalent to 6.1 Gaussian standard deviations,
thus confirming the observation of
$D^0$-$\bar{D}^0$ mixing.
 
\begin{acknowledgments}

  We thank the Fermilab staff and the technical staffs of the
  participating institutions for their vital contributions. This work
  was supported by the U.S. Department of Energy and National Science
  Foundation; the Italian Istituto Nazionale di Fisica Nucleare; the
  Ministry of Education, Culture, Sports, Science and Technology of
  Japan; the Natural Sciences and Engineering Research Council of
  Canada; the National Science Council of the Republic of China; the
  Swiss National Science Foundation; the A.P. Sloan Foundation; the
  Bundesministerium f\"ur Bildung und Forschung, Germany; the Korean
  World Class University Program, the National Research Foundation of
  Korea; the Science and Technology Facilities Council and the Royal
  Society, UK; the Russian Foundation for Basic Research; the
  Ministerio de Ciencia e Innovaci\'{o}n, and Programa
  Consolider-Ingenio 2010, Spain; the Slovak R\&D Agency; the Academy
  of Finland; the Australian Research Council (ARC); and the EU
  community Marie Curie Fellowship contract 302103.

\end{acknowledgments}

\bibliography{dmix_cdf.bib}   

\end{document}